\documentclass[10pt]{iopart}

\usepackage{graphicx}% Include figure files
\usepackage{dcolumn}% Align table columns on decimal point
\usepackage{bm}% bold math
\usepackage{subfigure}
\usepackage{multirow}
\expandafter\let\csname equation*\endcsname\relax
\expandafter\let\csname endequation*\endcsname\relax
\usepackage{amsmath}
\usepackage{mathrsfs}
\usepackage{mathtools}
\usepackage{comment}
\usepackage{algpseudocode}
\usepackage{relsize}
\usepackage{color}
\usepackage{amsmath}
\usepackage{xcolor}
\usepackage{cite} 
\usepackage{multicol}
\usepackage{hyperref}
\usepackage{cuted} % provides strip environment

\usepackage{xcolor}
         
\newcommand{\rt}[1]{\textcolor{black}{#1}}   
\newcommand{\rrt}[1]{\textcolor{black}{#1}}   
       
\usepackage{soul}
   
\usepackage[normalem]{ulem}
\newcommand{\rsmath}[1]{\bgroup\markoverwith{\textcolor{red}{\rule[0.5ex]{2pt}{0.4pt}}}\ULon {\textcolor{red}{#1}}}         
\usepackage{eso-pic}
\AddToShipoutPictureBG*{%
  \AtPageUpperLeft{%
    \hspace{\paperwidth}%
    \raisebox{-\baselineskip}{%
}}}%

\begin{document}

\paper{Designed self-assembly of programmable colloidal atom-electron equivalents}

\author{Xiuyang Xia$^{1,2}$\footnote[1]{These authors contributed equally to this work.}, Yuhan Peng$^1$\footnotemark[1], Ka Ki Li$^3$  and Ran Ni$^{1,\ast}$}

\address{$^1$ School of Chemistry, Chemical Engineering and Biotechnology,
Nanyang Technological University, 62 Nanyang Drive, 637459, Singapore}
\address{$^2$ Arnold Sommerfeld Center for Theoretical Physics and Center for NanoScience, Department of Physics, Ludwig-Maximilians-Universit\"at M\"unchen, Theresienstraße 37, D-80333 M\"unchen, Germany}
\address{$^3$ Department of Mathematics, Hong Kong University of Science and Technology, Clear Water Bay, Hong Kong}
\ead{\mailto{r.ni@ntu.edu.sg}}

\vspace{10pt}

\begin{abstract}
To unlock the potential for assembling complex colloidal ``molecules'', we investigate a minimal binary system of programmable colloidal atom–electron equivalents (PAE-EE), where electron equivalents (EEs) are multivalent linkers with two distinct types of single-stranded DNA (ssDNA) ends complementary to those ssDNAs on binary programmable atom equivalents (PAEs). We derive a statistical mechanical framework for calculating the effective interaction between PAEs mediated by EEs with arbitrary valency, which quantitatively agrees with simulations using explicit EEs. Our analysis reveals an anomalous dependence of PAE-PAE interactions on the EE valency, showing that EE-mediated interactions converge at the large valency limit. Moreover, we identify an optimal EE valency that maximizes the interaction difference between targeted and non-targeted binding pairs of PAEs. These findings offer design principles for targeted self-assembly in PAE-EE systems. 
%Our work provides tools for modeling soft matter systems with multivalent linkers, advancing the design of complex DNA structures and programmable materials.
\end{abstract}


\submitto{\RPP}
\maketitle
\ioptwocol

\section{Introduction}
DNA-based materials enable the self-assembly of complex and programmable structures through the precise molecular recognition capabilities of DNA~\cite{dnareviewscience}. By attaching single-stranded DNAs (ssDNAs) to colloidal surfaces, DNA-coated particles facilitate the formation of organized superstructures~\cite{mirkin1996dna,alivisatos1996organization,laramy2020crystal}, where the dual designability of DNA-encoded targeted structures and colloidal properties allows for the emulation of complex biological processes, merging biological sciences with material engineering, e.g., leading to applications in drug delivery~\cite{mura2013stimuli,hu2018dna}, biosensing~\cite{taton2000scanometric,cao2002nanoparticles,xia2024designing} and  immunotherapy~\cite{radovic2015immunomodulatory,wang2019rational}.

The linker-mediated scheme introduces additional designability for orthogonally regulating the specificity of pair interactions~\cite{biancaniello2005colloidal,xiong2009phase,zhang2015selective,lowensohn2019linker,xia2020linker,kwon2022dynamics}. 
Multivalent linkers, or termed electron equivalents (EEs), in these systems exhibit behavior analogous to electrons in metals, roaming through colloidal crystals and fostering effective attraction between larger DNA-coated colloids, termed programmable atom equivalents (PAEs)~\cite{girard2019particle}. This system serves as a colloidal analog to atoms and electrons, manifesting phenomena such as EE localization similar to the metal-insulator transition~\cite{girard2019particle,lin2020sublattice,ehlen2021metallization,lopez2021delocalization,lin2022superionic}, the emergence of coordination numbers~\cite{cheng2021electron,wang2022emergence}, and the formation of colloidal alloys~\cite{wang2019colloidal} and colloidal superionic conductors~\cite{lin2023colloidal}. 

Existing programmable atom-electron equivalent (PAE-EE) systems have been primarily utilizing EEs that are multivalent linkers with identical ssDNA ends. This homogeneity limits the complexity of the resulting assemblies. For more complex structures, such as colloidal ``molecules'', it is essential to employ EEs with heterogeneous ends to guide programmable assembly. \rt{Lowensohn et al.~\cite{lowensohn2019linker} and our previous work~\cite{xia2020linker} studied the system of linkers with heterogenous bi-valency, providing a foundation for understanding linker-mediated interactions. However, how the multivalency of EEs influences the phase behavior of the systems remains largely unexplored.} In particular, the design rule for selecting the optimal linker valency to achieve the best possible self-assembly remains unknown. \rt{Here, we extend the theoretical framework in Refs.~\cite{lowensohn2019linker,xia2020linker} and focus on the phenomena that emerge for \emph{arbitrary} EE valency.} We investigate a minimal binary PAE-EE system, where the EEs possess two distinct types of ends complementary to the ssDNA ends on binary PAEs. We derive an explicit formula for calculating the effective interaction between PAEs mediated by EEs of arbitrary valency from a statistical mechanical model of DNA coated colloids~\cite{angioletti2013communication,angioletti2014mobile,di2016communication,xia2020linker}. This approach allows us to study the distribution of EEs in different binding states, which can serve as a framework for Monte Carlo (MC) simulations with implicit EEs. We study the EE-mediated pair interaction between the same and different types of PAEs and corresponding phase behaviors, offering a design principle to guide the targeted self-assembly in PAE-EE systems.

\section{Methods}
 \subsection{Model}
\begin{figure}
 	\centering
 	\includegraphics[width=.47\textwidth]{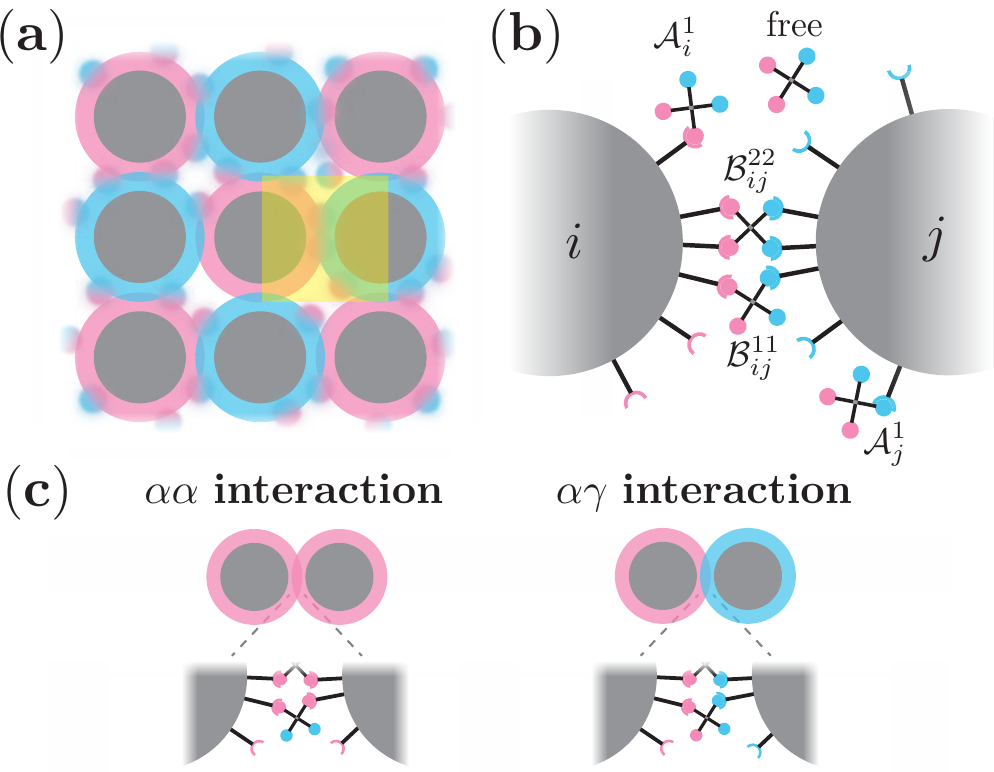}
  \caption{\label{fig1} \textbf{Programmable colloidal atom-electron equivalents.} (a) Schematic representation of binary PAEs, where the PAE-PAE interaction is mediated via the bridging of free EEs binding with ssDNAs on PAEs. (b) A magnification of the yellow region in (a). Mobile ssDNA ends (red or blue hooks) coated on PAEs can bind with the ends (red or blue spheres) of free EEs of the same color. EEs can stay in different binding substates characterized by the number of bound ends. (c) Schematic representation of two types of interactions in the system: non-targeted $\alpha\alpha$ interaction between the same type of PAEs and targeted $\alpha\gamma$ interaction between the different types of PAEs.}
 \end{figure}

We consider an \(\alpha \gamma\)-type binary PAE system in a solution containing free EEs, and each PAE is modeled as a hard sphere of radius \(R\) coated with \(n_{\rm \alpha}\) \(\alpha\)-type or \(n_{\rm \gamma}\) \(\gamma\)-type mobile ssDNA ends (Fig.~\ref{fig1}(a)). The mobility of ssDNAs coated on PAEs allows treating them as a mean-field-like adsorption layer to derive a closed form for effective interactions, and does not change the qualitative physics in the system~\cite{angioletti2014mobile}. ssDNA ends are modeled as infinitely thin, hard rods of length \(l_r\). 
Free EEs bridge PAEs in the solution and are connected to a reservoir of chemical potential \(\mu\).  Each EE is modeled as a polymer blob of zero radius, which is based on the fact that EEs are normally much smaller than PAEs experimentally. An EE consists of equimolar \(\lambda\) number of \(\alpha '\) and \(\gamma '\)-type ends, which are complementary to the $\alpha$  and $\gamma$-type ssDNAs coated on colloids, respectively.
We set the binding energy as \(\Delta G_{\rm \alpha \alpha '} = \Delta G_{\rm \gamma \gamma '} = \Delta G\), which characterizes the chemical details of specific binding. Non-specific bindings between \(\alpha-\gamma\), \(\alpha-\gamma'\), \(\alpha'-\gamma\), and \(\alpha'-\gamma'\) ends are neglected.
\subsection{EE-mediated interactions}
In typical experimental PAE-EE systems, van der Waals interactions and electrostatic interactions can be neglected~\cite{rogers2011direct,Varilly:2012gl}. Besides, we also neglect the depletion interaction originating from the EEs acting as depletants, particularly when EEs are sufficiently small. \rrt{Detailed justifications for these approximations are provided in the Supplementary Material Sec.~V.B~\cite{supinfo2}.} Based on the fact that PAEs are much larger than EEs~\cite{girard2019particle}, we assume that the motion of PAEs and EEs can be decoupled, and EEs move within the static external potential generated by PAEs and provide effective interactions among PAEs. Consequently, \rt{the effective interaction among PAEs can be expressed by}
\begin{equation}\label{eq:U_eff}
	\beta U_{\rm eff} = \beta F_{\rm att} - \beta F_{\rm att}' + \beta F_{\rm rep},
\end{equation}
\rt{which, when negative, indicates a favorable attractive interaction among the PAEs.} \(F_{\rm att}\) is the free energy of EEs given the coordinations of PAEs, \(F_{\rm att}'\) represents the free energy of EEs for isolated PAE configurations, \rt{and $\beta = 1/k_B T$ with $k_B$ and $T$ the Boltzmann constant and the temperature of the system, respectively}. The entropic repulsion \(\beta F_{\rm rep}\) originates from the excluded volume of ssDNAs overlapping with the hard cores of PAEs~\cite{angioletti2014mobile}.  \rt{We derive a statistical mechanical framework to calculate the effective interaction $\beta F_{\rm att}$ among PAEs mediated by EEs. To make the following derivations more accessible, we briefly outline the physical picture behind our statistical mechanical framework.}  As EEs are much smaller than PAEs, we assume that each EE can only be adsorbed onto one PAE $i$ (adsorbed state $\mathcal{A}_i$) or bridge between PAE $i$ and $j$ (bridging state $\mathcal{B}_{ij}$).  \rt{In a grand‑canonical description, the solution acts as a reservoir of EEs with a fixed chemical potential $\mu$, and the number of EEs in every possible binding substate is allowed to fluctuate.  We derive the corresponding partition function $\mathcal{Z}$ for the bound EEs on PAEs and, by applying the saddle‑point approximation, replace these fluctuating occupations by their most probable values. This yields closed‑form expressions for the average numbers of adsorbed and bridging EEs, $M_i$ and $Q_{ij}$.  Substituting $M_i$ and $Q_{ij}$ into the grand potential leads to an explicit and tractable formula for the EE‑mediated contribution $\beta F_{\rm{att}}$ that can be used directly in analytic calculations or Monte‑Carlo simulations.}

We note that EEs here can stay in different binding substates characterized by the number of bound ends. As shown in Fig.~\ref{fig1}(b), we denote $\mathcal{A}_{i}^{k}$ substate as $k$ ends on an EE binding with the ends on particle $i$, and $\mathcal{B}_{ij}^{kl}$ substate as $k$ and $l$ ends on an EE binding with the ends on PAE $i$ and $j$, respectively. \rt{The grand canonical partition function for the bound EEs on PAEs is (see Supplementary Material Sec.~II.A~\cite{supinfo2})
\begin{strip}
\begin{equation}\label{Eq:part_func_prim}
\begin{aligned}
    \mathcal{Z}\left(\{\mathbf{m}_i,\mathbf{q}_{ij}\}\right) = 
    \sum_{\{\mathbf{m}_i,\mathbf{q}_{ij}\}}
    \mathcal{W}\left(\{\mathbf{m}_i,\mathbf{q}_{ij}\}\right)
    \prod_i
    \left[\prod_{k}
    \left({\xi_{\mathcal{A},i}^{k}}\right)^{m_i^k}
    \prod_{j<i}\prod_{k,l}
    \left({\xi_{\mathcal{B},ij}^{kl}}\right)^{q_{ij}^{kl}}\right],
\end{aligned}
\end{equation}
\end{strip}
where $\mathbf{m}_i$ is a tuple of numbers ${m_i^k}$, where $k=1,\cdots,\lambda$ that represents the number of EE in $\mathcal{A}_i^k$ substate, and $\mathbf{q}_{ij}$ is a tuple of numbers $q_{ij}^{kl}$ that represents the number of EE in $\mathcal{B}_{ij}^{kl}$ substate. By counting the total number of ends in EE, we note that when $i$ and $j$ are the same types of particles, i.e., $t_i = t_j$, then for all $k+l\geq \lambda$, $q_{ij}^{kl}=0$.
$\{\mathbf{m}_i,\mathbf{q}_{ij}\}$ accounts for all possible values of $\mathbf{m}_i$ and $\mathbf{q}_{ij}$ for all particles and pairs of particles, respectively.  
The second and third terms represent the energy term for all EEs in adsorbed and bridging states, where $\prod_i \cdot$ indicates the product of a sequence on each particle $i$ from 1 to $N$, $\prod_{i,j<i}\cdot$ indicates the product of a sequence on each pair of particles $i$ and $j$, $\prod_{k} = \prod_{k=1}^{\lambda}$ denotes the product on each $\mathcal{A}_{i}^{k}$ adsorbed substate, and $\prod_{k,l}$ denotes the product on each $\mathcal{B}_{ij}^{kl}$ adsorbed substate with
\begin{equation}
	\prod_{k,l} = 
	\left\{
	\begin{aligned}
		\prod_{k=1}^{\lambda-1} \prod_{l=1}^{\lambda-k}, \quad t_{i} = t_{j}; \\
		\prod_{k=1}^{\lambda} \prod_{l=1}^{\lambda}, \quad t_{i} \neq t_{j}.
	\end{aligned}
	\right.
\end{equation}
Here, the individual partition function of EEs in substate $\mathcal{A}_{i}^{k}$ and $\mathcal{B}_{ij}^{kl}$ can be expressed as $\xi_{\mathcal{A},i}^{k} = m_i^0 (\chi_{i})^k$ and $\xi_{\mathcal{B},ij}^{kl} = m_{ij}^0 (\chi_{i})^k (\chi_{i})^l$, respectively.
$m_i^0 = V_{\mathcal{A},i} e^{\beta\mu}$ is the number of unbound EEs in the adsorption layer of PAE $i$, while $m_{ij}^0 = V_{\mathcal{B},ij} e^{\beta\mu}$ is the number of unbound EEs in the overlapped adsorption layers between PAE $i$ and $j$.}
\rt{Here $V_{\mathcal{A}, i}$ represents the remaining volume after excluding the hard sphere repulsion from neighboring particles and $V_{\mathcal{B},ij}$ represents the overlapping volume between the two adsorption layers from particle $i$ and $j$ because EE in a bridging state can only stay in this region.} \rt{$\chi_{i}= e^{-\beta\Delta G/\rho^{\circ} V_{\mathcal{A},i}}$ with \(\rho^{\circ}\) the standard concentration denotes the individual binding strength of each pair of ends binding between PAE $i$ and an EE, which accounts for both effects of energy and entropy (see Supplementary Material Sec.~II.D~\cite{supinfo2}). All possible combinations of PAE/EEs end binding can be expressed by (see Supplementary Material Sec.~II.E~\cite{supinfo2}):
\begin{equation}\label{Eq:Wall1}
\begin{aligned}
    \mathcal{W}\left(\{\mathbf{m}_i,\mathbf{q}_{ij}\}\right) = 
    & \prod_i \left[
    \frac{n_i!}{\bar{n}_i!}
    \prod_{k}
    \frac{\binom{\lambda}{k}^{m_i^k}}{m_i^k!}
    \prod_{j<i}\prod_{k,l}
    \frac{\left(w^{kl}\right)
    ^{q_{ij}^{kl}}}
    {q_{ij}^{kl}!} \right],  
 \end{aligned}
\end{equation}
where the combinatorial term in the bridged state
\begin{equation}
	w^{kl} = 
	\left\{
	\begin{aligned}
		\binom{\lambda}{k,l} &= \frac{\lambda!}{k!l!(\lambda-k-l)!}, \quad& t_{i} = t_{j}; \\
		\binom{\lambda}{k}\binom{\lambda}{l} &= \frac{\lambda!\lambda!}{k!l!(\lambda-k)!(\lambda-l)!}, \quad& t_{i} \neq t_{j}.
	\end{aligned}
	\right.
\end{equation}
and $\bar{n}_i$ denotes the number of unbound ends on particle $i$:
\begin{equation}\label{eq:constrain}
    \bar{n}_i = n_i - \sum_{k} k m_i^k - \sum_j \sum_{k,l} kq_{ij}^{kl}.
\end{equation}
}
Using the saddle-point approximation, we obtain the equilibrium number of EEs, ${m}_i^k$ and ${q}_{ij}^{kl}$, in different substates, $\mathcal{A}_{i}^{k}$ and $\mathcal{B}_{ij}^{kl}$, respectively (see \rt{Supplementary Material} Sec.~II.B~\cite{supinfo2})
\begin{equation}\label{eq:sadd_sol1} 
    {m}_i^k=(\bar{n}_i)^k \Xi_{\mathcal{A},i}^k, 
    \qquad
    {q}_{ij}^{kl}= (\bar{n}_i)^k(\bar{n}_j)^l \Xi_{\mathcal{B},ij}^{kl},
\end{equation}
where the binding strength of an EE in $\mathcal{A}_{i}^{k}$ substate, $\Xi_{\mathcal{A},i}^k$, and in $\mathcal{B}_{ij}^{kl}$ substate, $\Xi_{\mathcal{B},ij}^{kl}$, can be written as
\begin{subequations}\label{eq:binding_strength}
    \begin{align}
        \Xi_{\mathcal{A},i}^k&= m_i^0 \binom{\lambda}{k} (\chi_i)^{k},  
              \\
        \Xi_{\mathcal{B},ij}^{kl}&=m_{ij}^0  w^{kl}
         (\chi_{i})^k(\chi_{j})^l.
    \end{align}
\end{subequations}

Thus, the free energy of bound EEs is given by
\begin{equation}\label{eq:F_att}
\beta F_{\rm att}= \sum_i \left( n_i \log\bar{n}_i - \bar{n}_i - M_i -\sum_{j<i} Q_{ij} \right) + {\rm const},
\end{equation}
where $M_i$ and $Q_{ij}$ are the total numbers of EEs in the adsorbed state $\mathcal{A}_i$ and bridging state $\mathcal{B}_{ij}$, respectively, and the constant indicates the reference to different configurations of PAEs. \rt{Similarly, the free energy of EEs for isolated PAE configurations $\beta F_{\rm att}'$ is derived in Supplementary Material Sec.~II.C.} We notice that Eq.~\ref{eq:constrain} can be numerically solved, and $M_i$ and $Q_{ij}$ in Eq.~\ref{eq:F_att} are in closed forms for arbitrary $\lambda$ (see Supplementary Material  Sec.~II.B~\cite{supinfo2}). \rt{With the system consisting of $N$ particles, we obtain a set of $N$-element self-consistent equation with variables $(\bar{n}_1, \bar{n}_2, \cdots, \bar{n}_N)$ from Eq.~\ref{eq:constrain}, which can be numerically solved by fixed-point iteration. In Supplementary Material Sec.~II.F, we prove that there exists a unique solution $\{ \bar{n}_i \} $ for $0 \leq \bar{n}_i \leq n_i$~\cite{supinfo2}).} We note that when the EE is heterogeneously bivalent, i.e., $\lambda = 1$, the free energy can be simplified as
\begin{equation}
\begin{aligned}
\beta F_{\rm att}=
\sum_i \left[ n_i\log\frac{\bar{n}_i}{n_i} +\sum_{j<i} Q_{ij}
\right],
\end{aligned}
\end{equation}
which is consistent with the bivalent linker-mediated interaction expression~\cite{lowensohn2019linker,xia2020linker}.

The theory is based on the mean-field approximation, assuming negligible fluctuations in EE occupancy and rapid DNA binding/unbinding kinetics compared to colloidal dynamics. This approximation holds well when many EEs participate,  do not strongly interfere with each other, and binding equilibria can quickly adjust to changes in colloidal configurations. 
\rrt{Experimentally, EEs in assemblies are highly mobile, often exhibiting “metallic” behavior analogous to electrons in metals, where EEs roam freely throughout the colloidal crystal and reach equilibrium quickly~\cite{girard2019particle,lin2020sublattice}. This high mobility underlies the validity of the mean-field approach in our parameter regimes, as fluctuations are efficiently averaged out and correlations remain weak. Even in the transition to more ionic-like structures upon crystallization, the delocalization-to-localization of EEs occurs continuously rather than abruptly~\cite{lopez2021delocalization}, suggesting that the mean-field description remains appropriate until EEs become strongly localized or comparable in size to the PAE unit cell. Moreover, we do not explicitly include crowding among EE strands, as their concentrations are low and each EE typically carries only a few DNA arms under experimental conditions, making the steric penalty much smaller than for densely grafted PAE receptors. If necessary, excluded volume effects between EEs could be incorporated via an additional chemical-potential correction to the free energy, and this would introduce extra parameters without affecting the essential physics of multivalency ~\cite{Hagan2004}.}

\section{Results}
 \subsection{Numerical verification}
\begin{figure}
 	\centering
 	\includegraphics[width=.47\textwidth]{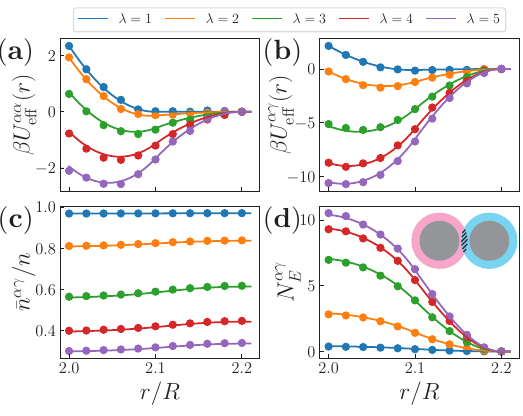}
 	\caption{\label{fig2} \textbf{Numerical verification of the effective pair interactions between PAEs.} (a,b) EE-mediated effective PAE-PAE interactions $\beta U_{\rm eff} (r)$ as functions of the center-to-center distance $r$ between two PAEs for different EE valency $\lambda$, where the PAEs are of the same type (a, $\beta U_{\rm eff}^{\alpha\alpha}$) and different types (b, $\beta U_{\rm eff}^{\alpha\gamma}$). \rt{Note a negative effective potential $\beta U_{\mathrm{eff}}$ corresponds to an attractive interaction.} (c) The ratio of unbound ssDNAs $\bar{n}^{\alpha\gamma}/n$ as a function of $r$ for different $\lambda$ between PAEs of different types. (d) Number of EEs $N_{\rm E}^{\alpha\gamma}$ depicted by the shadow area in the inset as a function of $r$ for different $\lambda$ between PAEs of different types. Symbols are from numerical simulations, and curves are theoretical predictions. 
 	Parameters: $\beta\mu = -14$, $\beta \Delta G =-10$, $n=100$ and $l_r = 0.1 R$.}
 \end{figure}

We aim to guide the targeted binding between $\alpha$ and $\gamma$ type PAEs through EE-mediated interactions. However, as shown in Fig.~\ref{fig1}(c), due to the multivalency of EEs, \rt{EE-mediated interactions exist between particles of the same type and different types. We consider $\beta U_{\rm eff}^{\alpha \gamma/ \alpha \alpha}$ to be the targeted/non-targeted PAE-PAE interaction, respectively.} Our goal is to understand how to design these interactions to optimize the ability to form targeted structures. We first perform grand canonical Monte Carlo (GCMC) simulations incorporating explicit EEs to validate the pair interactions. In our simulations, PAEs are modeled as mobile DNA-coated colloids (mDNACCs), while the EEs are treated as zero-volume mDNACCs that can interact with both $\alpha$ and $\gamma$ PAEs. The interactions within the system are calculated based on the framework of reversible mobile binders proposed in Ref.~\cite{angioletti2014mobile}. Two PAEs of the same ($\alpha\alpha$) or different types ($\alpha\gamma$) are placed at fixed positions within a simulation box, while the EEs are inserted and deleted using standard GCMC moves~\cite{frenkel2023understanding}. The effective interactions are calculated for a given chemical potential of EEs $\mu$ and a specified center-to-center distance $r$ between two PAEs using the thermodynamic integration (see \rt{Supplementary Material} Sec.~IV.A~\cite{supinfo2}). As shown in Fig.~\ref{fig2} (a,b), we plot the pair interactions between the same and different types of PAEs, $\beta U_{\rm eff}^{\alpha\alpha}$ and $\beta U_{\rm eff}^{\alpha\gamma}$, demonstrating that our method quantitatively agrees with the explicit EE simulations without any fitting parameter. Our method captures the number of unbound ssDNAs on PAEs, i.e., $\bar{n}^{\alpha \gamma} / n$ with $n_{\alpha} = n_{\gamma} = n = 100$, and the number of EEs between two PAEs, i.e., $N_{E}^{\alpha \gamma}$, and these results also agree quantitatively with the simulation results (Fig.~\ref{fig2}(c,d)). Furthermore, our method provides detailed information on EEs in different substates.
 
\begin{figure*}
 	\centering
\includegraphics[width=1\textwidth]{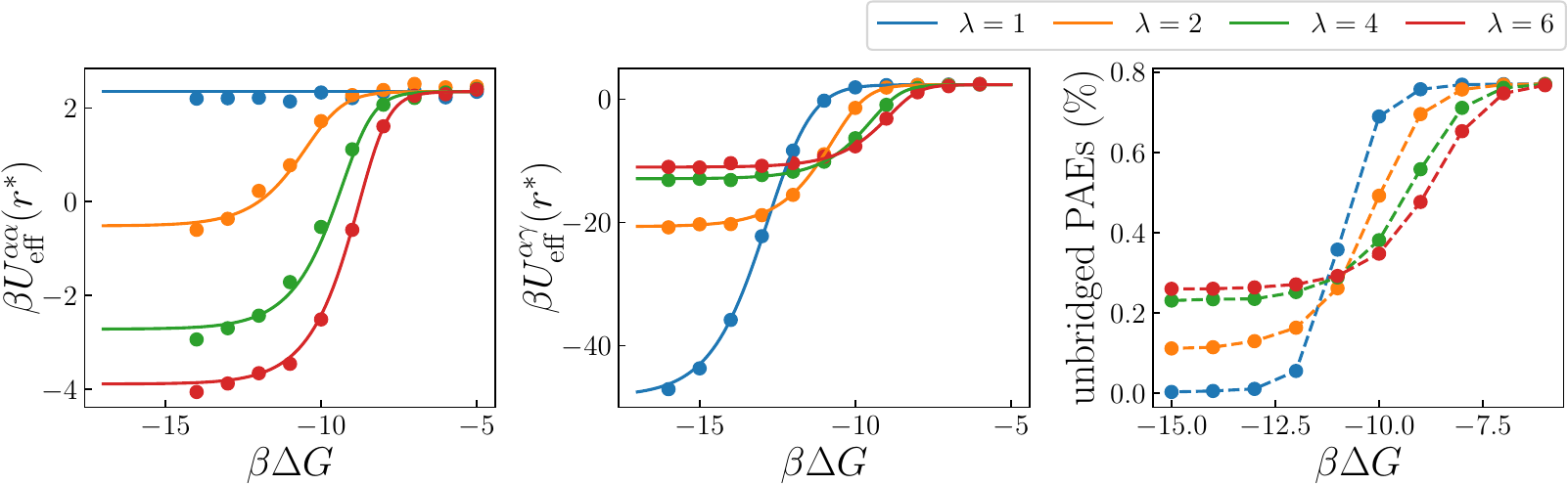}
 	\caption{{\label{fig3} \textbf{Entropy driven PAE-PAE interactions\rt{.}} (a,b) The strength of EE-mediated interaction $\beta U_{\rm eff}(r^*)$ as functions of individual binding energy $\beta \Delta G$ between two PAEs for different EE valency $\lambda$ at $\beta\mu = -13$, where the PAEs are of the same type (a, $\beta U_{\rm eff}^{\alpha\alpha} (r^*)$) and different types (b, $\beta U_{\rm eff}^{\alpha\gamma} (r^*)$). Symbols are from numerical simulations, and curves are theoretical predictions. (c) \rt{percentage of unbridged PAEs} in manybody PAE-EEs systems at $\eta = 0.094$ as a function of $\beta \Delta G$ for different $\lambda$ at $\beta\mu = -11$. \rt{Symbols are obtained by the many-body Monte Carlo simulations, and the dashed lines are guided by eyes.} Parameters: $n_{\alpha} = n_{\gamma} = 100$,  $r^\ast = 2R$  and $l_r = 0.1 R$.  }}
\end{figure*} 
\subsection{Entropy driven effective PAE-PAE interactions}
As shown in Fig.~\ref{fig3}(a,b), similar to the bivalent linker-mediated DNACCs and polymer-microemulsion networks~\cite{xia2020linker,zilman2003entropic},  we observe that the strength of effective colloidal interactions at contact $\beta U_{\rm eff}^{\alpha \alpha/ \alpha \gamma}(r^\ast)$ \rt{saturates and does not diverge at the strong binding or low temperature limit under any valency of EEs}. The reason is that at \rt{the strong binding} limit, all ssDNAs on any  PAE $i$ are occupied, i.e.,  $\bar{n}_i\to 0$, therefore, the energy remains the same as the configuration where all PAEs are isolated for any PAE configurations. \textcolor{black}{As shown in Eq.~\ref{eq:U_eff}, the energy term cancels out in $\beta F_{\rm att}$ and $\beta F_{\rm att}'$ and therefore does not contribute to the effective interaction $\beta U_{\rm eff}$.} At this limit, $\beta U_{\rm eff}$ solely depends on the combinatorial and configurational entropies, i.e., the valency and chemical potential of EEs. Based on the additional constraint $\bar{n}_i \to 0$ and taking $\mathcal{A}^1$ as the reference state, we introduce the entropy driven effective binding strength of an EEs under $\mathcal{A}_{i}^{k}$ substate, $\tilde{\Xi}_{\mathcal{A},i}^k$, and under $\mathcal{B}_{ij}^{kl}$ substate, $\tilde{\Xi}_{\mathcal{B},ij}^{kl}$, respectively (see \rt{Supplementary Material} Sec.~III.A~\cite{supinfo2})
\begin{subequations}
\begin{align} 
\tilde{\Xi}_{\mathcal{A},i}^k =& \left. \frac{\Xi_{\mathcal{A},i}^k}  {\left(\Xi_{\mathcal{A},i}^1\right)^k} \right.
	 =  \frac{\binom{\lambda}{k}  m_i^0}{\left(  m_i^0 \lambda \right)^k}, \\
\tilde{\Xi}_{\mathcal{B},ij}^{kl} =&  \left. \frac{\Xi_{\mathcal{B},ij}^{kl}}  {\left[{\left(\Xi_{\mathcal{A},i}^1\right)^k \left(\Xi_{\mathcal{A},j}^1\right)^l}\right]} \right.
    =     \frac{w^{kl}m_{ij}^0}{ \left(  m_i^0 \lambda \right)^k \left(  m_j^0 \lambda \right)^l},
\end{align}
\end{subequations}
with the free energy of bound EEs 
\begin{equation}\label{eq:fatt_limit}
\begin{aligned}
\beta F_{\rm att}=
\sum_i \left[ n_i \log \left (\frac{m_i^1}{ m_i^0} \right) - M_i -\sum_{j<i}Q_{ij} \right] + \mathrm{const.}
\end{aligned}
\end{equation}
To demonstrate the entropy driven PAE-PAE interactions in many-body systems, we perform $NVT$ MC simulations with the effective interaction from Eq.~\ref{eq:U_eff} and \ref{eq:F_att} for an equimolar binary mixture of PAEs at the packing fraction $\eta = 4N\pi R^3 / 3V = 0.094$ with various $\beta \Delta G $ and $\lambda$ (see \rt{Supplementary Material} Sec.~IV.B~\cite{supinfo2}). As shown in Fig.~\ref{fig3}(c), the \rt{percentage of unbridged PAEs } decreases and reaches a plateau with decreasing $\beta \Delta G$, following the same trend as the pair interaction (Fig~\ref{fig3}(a,b)). \rt{Additional many-body simulations at higher packing fractions ($\eta =$ 0.094–0.282, Fig. S1) show the same unbridged-fraction trends, confirming that our conclusions are insensitive to the PAE concentration.}
  \begin{figure}
 	\centering
 	\includegraphics[width=.47\textwidth]{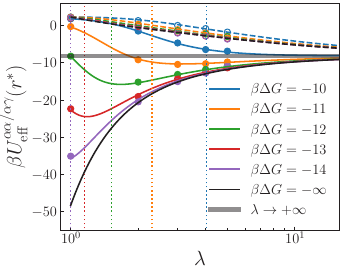}
 	\caption{\label{fig4}\textbf{Anomalous $\lambda$ dependence of effective interaction.}  The strength of EE-mediated interactions between PAEs of the same type ($\beta U_{\rm eff}^{\alpha\alpha}$, dashed curves and open symbols) and the different types ($\beta U_{\rm eff}^{\alpha\gamma}$, solid curves and solid symbols) as functions of EE valency $\lambda$ with different binding energy $\beta \Delta G$. Symbols are calculated from numerical simulations. The vertical dotted lines show the location of optimal valency $\lambda^*$ where $|\beta U_{\rm eff}^{\alpha\gamma} (r^*) - \beta U_{\rm eff}^{\alpha\alpha} (r^*)|$ maximizes at different $\beta \Delta G$. The horizontal grey line indicates $\beta U_{\rm eff} (r^*)$ at large $\lambda$ limit. Parameters: $\beta\mu = -13$, $n = 100$,  $r^\ast = 2R$  and $l_r = 0.1 R$.
       }
 \end{figure}
 \subsection{Anomalous EE valency dependence}
As shown in Fig.~\ref{fig3}(a,b), approaching the strong binding limit ($\beta \Delta G \to -\infty$), intriguingly, the effective interaction strength at contact between the same type of PAEs, $\beta U_{\rm eff}^{\alpha \alpha}(r^*)$, becomes stronger as the valency of EE $\lambda$ increases, while the effective interaction at contact between different types of PAEs, $\beta U_{\rm eff}^{\alpha\gamma}(r^*)$, shows the opposite dependence. 
\rt{Numerical simulations suggest that the trends also hold in the case when EEs are of finite volume (Fig. S2).}
To clarify these trends, we plot $\beta U_{\rm eff}^{\alpha\alpha/\alpha\gamma}(r^*)$ as functions of $\lambda$ at the strong binding limit according to Eqs.~\ref{eq:U_eff} and \ref{eq:fatt_limit}, shown as the black solid and dashed curves in Fig.~\ref{fig4}.
We note that at this limit, for an isolated PAE $i$, the number of EEs that stay in adsorbed states $M_i$ decreases as EE valency increases since fewer EEs are enough to saturate the receptors on the PAE. Moreover, the effective pair interaction in the system mainly depends on $Q$ with $\beta U_{\rm eff} \approx -Q$. 
For the $\alpha\gamma$ interaction, the bridging of EEs is determined by the adsorption on both particles, where adsorption promotes bridging with $Q_{ij} \sim M_i M_j$. Therefore, as $\lambda$ increases, both the numbers of EEs in the adsorbed state and bridging state decrease, leading to a weaker effective EE-mediated interaction. 
Differently, for the $\alpha\alpha$ interaction, we have $Q_{ij} \sim -M_i - M_j$, indicating that bridging competes with adsorption. Thus, increasing the $\lambda$ decreases the number of EEs in the adsorbed state while increasing the number of EEs in the bridging state, resulting in a stronger effective PAE interaction. At finite binding energies, the effect of increasing \(\lambda\) also depends on the types of PAEs involved. As shown in Fig.~\ref{fig4}, for \(\alpha\alpha\) interactions, increasing the EE valency leads to stronger interactions that eventually converge, similar to the behavior in the strong binding limit. In contrast, the situation is more complex for \(\alpha\gamma\) interactions. At relatively weak binding energies (e.g., \(\beta \Delta G = -10\)), increasing \(\lambda\) strengthens the PAE-PAE interaction; at strong binding energies (e.g., \(\beta \Delta G = -14\)), increasing the EE valency weakens the PAE-PAE interaction; and at intermediate binding energies (e.g., \(\beta \Delta G = -12\)), the dependence of the strength of the PAE-PAE interaction on \(\lambda\) is non-monotonic.
The reason for this behavior is that as the binding energy becomes stronger or the valency becomes larger, receptors on the PAEs are more easily saturated. Consequently, the effective interaction \(\beta U_{\mathrm{eff}}^{\alpha\gamma}(r^*)\) more closely approaches that of the strong binding limit.

Furthermore, interestingly, we observe that with increasing $\lambda$, indicated by the gray horizontal line in Fig.~\ref{fig4}, the effective PAE interactions unexpectedly converge and become independent of the binding energy and the type of pairs. At this limit, we also have $\bar{n}_i \to 0$, similar to the strong binding limit according to Eq.~\ref{eq:binding_strength}. Thus, $\beta F_{\rm att}$ in Eq.~\ref{eq:fatt_limit} holds with the effective binding strength (see \rt{Supplementary Material} Sec.~III.B~\cite{supinfo2})
\begin{equation} 
\tilde{\Xi}_{\mathcal{A},i}^k = \frac{m_i^0}{k! {\left(m_i^0\right)}^{k}},\qquad \tilde{\Xi}_{\mathcal{B},ij}^{kl}= \frac{m_{ij}^0} {k!l! \left(m_i^0\right)^{k} \left(m_j^0\right)^{l}}. 
\end{equation} 
This indicates that increasing the valency has less and less impact on effective interactions. 

As all interactions converge at large $\lambda$, one can find that there exists an optimal valency $\lambda^*$ at which the difference between strengths of effective interactions among different PAEs  $|\beta U_{\rm eff}^{\alpha\gamma}(r^*) - \beta U_{\rm eff}^{\alpha\alpha}(r^*)|$ maximizes, and $\lambda^*$ decreases with decreasing the binding free energy $\Delta G$ (Fig.~\ref{fig4}). This optimal $\lambda$ maximizes the disparity between targeted ($\alpha  \gamma$) and non-targeted ($\alpha\alpha$) PAE interactions, and our results suggest that at the strong binding limit ($\Delta G \to -\infty$), the optimal valency of EE for mediating specific binding between targeted PAEs is $\lambda^* = 1$. However, we note that in experiments of designed assembly, the binding strength is normally not very strong, as strong binding would make it hard for the system to reach the designed state in equilibrium. \rt{We note that up to 10\% polydispersity in the number of grafted strands shows less than $ 0.3 k_BT$ deviations in the pair potentials and leaves the optimal valency $\lambda^{*}$ essentially unchanged (Fig. S3), indicating that our design rules are robust to the level of polydispersity reported experimentally.}

\section{Discussion}
To conclude, we have studied the self-assembly of a minimal binary system of PAE-EEs, where EEs act as multivalent linkers with two types of ssDNA ends complementary to those on PAEs. We derived an explicit formula for the effective PAE-PAE interaction mediated by EEs with arbitrary valency $\lambda$, which quantitatively agrees with simulations incorporating explicit EEs. \rt{The formulas can be obtained by generalizing the approach in Ref~\cite{lowensohn2019linker,xia2020linker} to linkers whose multivalency $\lambda>1$. Here we focus on how the multivalency of linkers influences the effective interaction and phase behaviors in the system.} Our analysis reveals an anomalous dependence of the effective PAE-PAE interactions on $\lambda$. Specifically, at the strong binding limit, we found that increasing $\lambda$ strengthens the interaction between PAEs of the same type but weakens that between different types, which converges to the same value in the large valency limit. We identified an optimal valency $\lambda^*$ that maximizes the effective interaction difference between the targeted and non-targeted PAE pairs, which provides a practical design principle for the designed self-assembly of PAE-EE systems. \rt{The design principle can be implemented directly with current DNA-nanotechnology protocols.  Orthogonal 6–8 mer sticky ends that satisfy the required $\Delta G$ can be generated with standard sequence–design software. EE valency is controlled experimentally by varying the density of thiolated oligonucleotides during functionalisation~\cite{wang2022emergence}.} Moreover, our explicit formula can be directly used in MC simulation to efficiently sample the thermodynamic and structural information of the system. \rt{The same saddle-point scheme can be applied to ternary or more complex mixtures by enlarging the binding matrix, and no additional physics is required.} 

Beyond motivating experimental work towards the design of complex DNA structures, we provide the tools for modeling and analyzing soft matter and biological systems with multivalent linkers, e.g., polyelectrolyte~\cite{muthukumar201750th,yu2018multivalent}, hydrogels~\cite{ooi2020multivalency,le2024valence}, vitrimers\cite{rottger2017high, lei2020entropy}, multivalent drug design~\cite{mammen1998polyvalent,dubacheva2023determinants}, and cell adhesion and signaling~\cite{zhu2008structure}. 
\rt{While our receptor-mobility model already includes many-body effects among PAEs, higher-order bridges in which a single EE binds three or more particles are not considered here and will be addressed in future work.}
\rt{Moreover, kinetic pathways or possible arrest often accompany short-range attractive colloids~\cite{Jack2007,Klotsa2011,Haxton2015}, and coupling our multivalent-linker framework to real-time kinetic analyses is a promising avenue for future study.}

\section*{Data availability statement}
All data that support the findings of this study are included within the article (and any supplementary files). \rt{All scripts used to compute the PAE–EE pair interactions and to run the many-body Monte Carlo simulations are openly available under the MIT licence at \url{https://github.com/xiaxiuyang/multivalent_linker_mediated_interaction}.}

\ack{This work was financially supported by the Academic Research Fund from the Singapore Ministry of Education (RG151/23 and MOE2019-T2-2-010) and the National Research Foundation, Singapore, under its 29th Competitive Research Program (CRP) Call (NRF-CRP29-2022-0002). X. X. acknowledges support from the Alexander von Humboldt-Stiftung.}

\section*{References}
\bibliographystyle{iopart-num} 
\bibliography{abbr}           

\end{document}